\documentclass[pra,amsfonts,tightenlines,twocolumn,showpacs,superscriptaddress,12pt]{revtex4}%this is for the final version
\usepackage{amsmath}
\usepackage{amsthm}
\input{psfig.sty}

\newcommand\set[1]{\ensuremath{\{#1\}}}
\newcommand\smatrixx[1]{\left(\begin{smallmatrix}#1\end{smallmatrix}\right)}

\def\opone{\leavevmode\hbox{\small1\kern-3.8pt\normalsize1}}
\newtheorem{theo}{Theorem}
\theoremstyle{remark}
\newtheorem*{remark}{Technical remark}
\renewcommand{\O}{\textrm{O}}
\newcommand{\RR}{\mathbb{R}}
\newcommand{\RRR}{\mathbf{R}}
\newcommand{\CC}{\mathbb{C}}
\newcommand{\NN}{\mathbb{N}}
\newcommand{\ZZ}{\mathbb{Z}}
\newcommand{\PP}{\mathbb{P}}
\newcommand{\XX}{\mathbf{X}}
\newcommand{\YY}{\mathbf{Y}}
\newcommand{\q}{\quad}
\newcommand{\KK}{\mathbb{K}}
\newcommand{\ee}{\mathbf{e}}

\newcommand{\eps}{\boldsymbol{\epsilon}}

\begin{document}
\title {Weak limits for quantum random walks}

\author{Geoffrey Grimmett}
\email{g.r.grimmett@statslab.cam.ac.uk}
\affiliation{Statistical Laboratory, University of Cambridge,
Centre for Mathematical Sciences,
Wilberforce Road, Cambridge CB3 0WB, UK}

\author{Svante Janson}
\email{svante.janson@math.uu.se}
\affiliation{Department of Mathematics, Uppsala University, PO Box 480,
S-751 06 Uppsala, Sweden}

\author{Petra F. Scudo}
\email{p.scudo@statslab.cam.ac.uk} \affiliation{Statistical
Laboratory, University of Cambridge, Centre for Mathematical
Sciences, Wilberforce Road, Cambridge CB3 0WB, UK}
\affiliation{Department of Physics, Technion -- Israel Institute of
Technology, 32000 Haifa, Israel}

\begin{abstract}
%\textbf{Abstract.}
We formulate and prove a general weak
limit theorem for quantum random walks in one and more dimensions.
With $X_n$ denoting position at time $n$, we show
that $X_n/n$ converges weakly as $n\to \infty$ to a certain
distribution which is absolutely continuous and of bounded support.
The proof is rigorous and makes
use of Fourier transform methods.
This approach simplifies and extends certain preceding
derivations valid in one dimension that make use
of combinatorial and path integral methods.
\end{abstract}
\maketitle

\section{Introduction}
Let $R_1,R_2,\dots$ be independent identically
distributed random
variables taking values in the reals $\RR$, and suppose that
they have common mean $\mu=E(R_1)$ and finite
non-zero variance $\sigma^2=E(R_1^2)-\mu^2$. The central limit
theorem asserts that the sum $X_n=\sum_{i=1}^n R_i$ satisfies
\begin{equation}
\frac{X_n-n\mu}{\sigma\sqrt n} \Rightarrow \textrm{N}\q \text{as}\
n\to\infty \label{clt}
\end{equation}
where N denotes the normal (Gaussian) distribution with mean $0$
and variance $1$, and $\Rightarrow$ denotes weak convergence:
\begin{equation}
T_n\Rightarrow T \q \text{if} \q E(f(T_n))\to E(f(T))
\end{equation}
for all bounded continuous functions $f: \RR\to\RR$.
An early version of this now classical theorem for random walks
was proved as long
ago as 1733 by de Moivre, \cite{grimmett}.
In the modern theory, the conditions on the $R_i$ are relaxed to allow
non-independent non-identically distributed random variables taking
values in general spaces. Since the weak limit of $X_n$, suitably normalized,
depends only on the probability measures associated with the $X_n$,
we may think of the central limit theorem as a result about weak limits
of measures, rather than about the stochastic process $(X_n:n\ge 1)$ itself.
This is an impoverishment of the theory, since it overlooks
the random variables themselves.

There has been recent interest, \cite{kempe}, in a new type of
process termed a quantum random walk. Quantum random walks give
rise to certain sequences $(\mu_n:n\ge 1)$ of probability
measures, each of which is given in terms of the preceding
measures in the sequence. Whilst it is possible as always to
construct random variables having these measures, this may not be
done in a natural manner as in the theory of stochastic processes.
One may nevertheless ask whether, subject to an appropriate
normalization, the $\mu_n$ converge weakly to some non-trivial
distributional limit.  Results in this direction have been
obtained for one-dimensional quantum walks by Konno \cite{konno1,
konno2}. We show in this note how to simplify and extend such
results. We introduce a new method of studying such weak limits,
and we apply this method to quantum walks in one and higher
dimensions.

We consider first a quantum random walk on the integers $\ZZ$. At
each time $n$ ($\in \NN$) the state of the particle
is transformed by a unitary operator described by a rotation of
the internal degree of freedom followed by a conditional shift of
the position, \cite{nayak};
the internal degree of freedom represents a coin that determines the
shift of the position.
 The overall state of the system belongs
to the Hilbert space $H_C \otimes H_P$, where $H_C$ is associated
with the internal degree of freedom (coin space) and $H_P$ with
position. In the simplest case, we have $H_C=\CC^2$ and
$H_P=\ell^2(\ZZ)$. A suitable basis for $H_P$ is given by
the eigenstates of the position operator $X$
\begin{equation}
X  v_x = x   v_x, \q\q x\in\ZZ,
\end{equation}
subject to $\langle v_x, v_{x'}\rangle=
\delta_{xx'}$, the Kronecker delta.
A general state of the system may be
written with respect to this basis as
\begin{equation}
\psi=\sum_x \sum_j \psi_j (x)  v_x w_j,
\end{equation}
where the vectors $w_j$, $j= 1, 2$, define a standard basis in
$H_C$. The probability $\mu_n(x)$ of finding the particle at the position $x$
at time $n$ is given by the standard rule
\begin{equation}
\mu_n(x)=\sum_j|\langle v_x \, w_j, \psi_n\rangle|^2 \label{prob}
\end{equation}
where $\psi_n = U^n \psi_0$ with $U$ the time-evolution operator of
the walk and $\psi_0$ the initial state of the system.

The asymptotic properties of the sequence $(\mu_n:n\ge 1)$
are studied in the next section. Such results are extended in Section III
to quantum walks in two and more dimensions. We highlight two
special features of such asymptotics, namely:
instead of normalizing by $\sqrt n$ as in (\ref{clt}),
we shall normalize by $n$,
and the weak limit is absolutely continuous with bounded support.

\section{Weak limit for one-dimensional quantum walks}

In order to define the position of a quantum particle as a random
variable, we consider the evolution of the position operator in the
Heisenberg picture starting from time $n = 0$. At each time
$n$, the eigenvalues of the operator $X_n \doteq U^{\dag n} X
U^{n}$ define the possible values of the particle's position
with corresponding probability given by (\ref{prob}), where the
dependence
on the initial state $\psi_0$ is explicit.

Although the position may be treated as an ordinary random
variable, the sequence $(X_n:n\ge 1)$ does not define a stochastic process,
since the simultaneous measurement of $X_n$ for different $n$
would change the quantum
random walk at each step. Therefore we let the system evolve
repeatedly under $U$ up to time $n$, without measuring it, and
then we study the properties of the distribution $\mu_n$ of $X_n$.

Let $\psi_0$ be any initial state in $H_C \otimes H_P$ with all
moments $E (X^r)$ finite. In order to simplify the calculations
which follow, we consider
transformations in terms of wave function components, and we take the
Fourier transform space $\widehat{\ell^2(\ZZ)}=
L^2(\KK)$, where $\KK=[0,2\pi)$ is thought of
as the unit circle in $\RR^2$.
We define an inner product on $L^2(\KK)$
by
\begin{equation}
\langle\psi,\phi\rangle=\int_{0}^{2\pi}
\overline{\psi(k)}\phi(k)\,\frac{dk}{2\pi}
\end{equation}
and we note the isometry  between
$\ell^2(\ZZ)$ and $L^2(\KK)$ given  by
\begin{equation}
(\psi_x)\mapsto\sum_x \psi_x e^{ixk},
\end{equation}
with inverse
\begin{equation}
 \psi\mapsto \hat \psi\q\text{where}\q  \hat
\psi(x)=\int_{0}^{2\pi}e^{-ixk} \psi (k) \,\frac{dk}{2\pi}.
\end{equation}
The right shift $S$ on
$\ell^2(\ZZ)$ given by
$S(\psi_x)_{-\infty}^\infty=(\psi_{x-1})_{-\infty}^\infty$
corresponds to the multiplication operator $\hat S \psi = e^{ik}
\psi$ on $L^2(\KK)$.

Our fundamental Hilbert space is thus $H=H_C\otimes
L^2(\KK)$, the space of
$\CC^2$-valued functions
\begin{equation}
\psi(k)=\left ( \begin{array}{c}\psi_1(k)\\
\psi_2(k)\end{array}\right)
\end{equation}
on $\KK$ satisfying
$$
\|\psi\|^2=\|\psi_1\|_{L^2}^2+\|\psi_2\|_{L^2}^2 <\infty.
$$
As usual, we consider state vectors normalized by
$
\|\psi\|^2= 1
$.
The
evolution of the walk comprises repeated applications of an
internal transformation (coin toss) $A$ acting on $\CC^2$,
followed by the shift $S$ given by
\begin{equation}\label{s11}
S \left ( \begin{array}{c}\psi_1(k)\\
\psi_2(k)\end{array}\right)
 = \left ( \begin{array}{c}e^{ik}\psi_1(k)\\
e^{-ik} \psi_2(k)\end{array}\right).
\end{equation}
Thus the total evolution $U$ on $H$ is given by
\begin{equation}\label{uk}
U \psi= \left( \begin{array}{cc} e^{ik} & 0 \\ 0 & e^{-ik}\end{array} \right)
A \left ( \begin{array}{c}\psi_1(k)\\
\psi_2(k)\end{array}\right)
 =U(k)\psi(k).
\end{equation}
If we begin the quantum random walk with an initial state $\Psi_0
\in H$, its state after $n$ steps is
\begin{equation}
\Psi_n=
U^n\Psi_0=U(k)^n\Psi_0(k).
\end{equation}

For each $k$, $U(k)$ has two eigenvalues $\lambda_1(k)$ and
$\lambda_2(k)$ with $|\lambda_j(k)|=1$, and has corresponding
eigenvectors $v_1(k),\,v_2(k)\in \CC^2$ that define a
basis for $H$. We assume henceforth that
\begin{equation}
\lambda_1(k)\neq
\lambda_2(k), \label{assump}
\end{equation}
since otherwise $U(k)$ is diagonal; by \eqref{uk} then $A$ is diagonal
and the state
evolves trivially, either to the right or to the left.

The mapping $k\mapsto U(k)$ is $C^\infty$ and the eigenvalues are
distinct for each $k$, and therefore the eigenvalues $\lambda_j(k)$ are
$C^\infty$ functions of $k$, and the eigenvectors $v_j(k)$ may be
chosen to be $C^\infty$ with normalization $\|v_j(k)\|=1$.
By expanding the wave function in terms of this basis, the $n$th
time evolution becomes
\begin{widetext}
\begin{equation}
\label{a} \Psi_n(k)=U(k)^n\Psi_0(k) =\lambda_1(k)^n\langle
v_1(k),\Psi_0(k) \rangle v_1(k) + \lambda_2(k)^n\langle
v_2(k),\Psi_0(k)\rangle v_2(k),
\end{equation}
\end{widetext}
where each component on the right hand side is a $C^\infty$
function of $k$. The moments of the position distribution are
given in terms of the operator $X$ according to the standard formula
\begin{equation}
\label{b1} E \left(X_n^r\right) = \langle \Psi_n,X^r\Psi_n\rangle.
\end{equation}

Using the isometry between $\ell^2(\ZZ)$ and
$L^2(\KK)$, the above expectation may be written as
\begin{equation}
\label{b2}  E \left(X_n^r\right) = \int_{0}^{2 \pi}\langle
\Psi_n(k),D^r\Psi_n(k)\rangle \,\frac{dk}{2\pi},
\end{equation}
where $D = \hat{X}= -i {d}/{dk}$ is the position operator in
the momentum space $L^2(\KK)$. For fixed $r$ we can compute
$D^r\Psi_n(k)$ by (\ref{a}) and Leibniz' rule. It is easily seen
that
\begin{align}
\label{c} D^r\Psi_n(k) =&\sum_j (n)_r \lambda_j(k)^{n-r}
(D\lambda_j(k))^r\nonumber\\
&\times\langle v_j(k),\Psi_0(k)\rangle v_j(k)
+\O(n^{r-1}),
\end{align}
where  $(n)_r=n(n-1)\cdots(n-r+1)$.
Equations (\ref{b2}) and (\ref{c}) yield, as $n\rightarrow\infty$,
\begin{widetext}
\begin{align}
E [(X_n/n)^r]
&= \int_0^{2\pi}\sum_j \lambda_j(k)^{n-r}
\left(D\lambda_j(k)\right)^r \langle v_j(k),\Psi_0(k)\rangle
\langle \Psi_n(k),v_j(k)\rangle\,\frac{dk}{2\pi} +\O(n^{-1})\nonumber \\
&\label{d}=\int_{0}^{2\pi}\sum_j
\left(\frac{D\lambda_j(k)}{\lambda_j(k)}\right)^r |\langle
v_j(k),\Psi_0(k)\rangle|^2 \,\frac{dk}{2\pi} +\O(n^{-1}).
\end{align}
\end{widetext}

Let $\Omega = \KK\times \{1,2\}$, let $\mu$ be
the probability measure on $\Omega$ given by $|\langle
\Psi_0(k),v_j(k)\rangle|^2 dk/2\pi$ on $\KK\times\{j\}$.
Let $h_j(k)=\lambda_j(k)^{-1}D\lambda_j(k)$ and define
$h:\Omega
\rightarrow \RR$ by $h(k,j)=h_j(k)$.
($h$ is real because $|\lambda_j(k)|=1$.)
By (\ref{d}),
\begin{equation}
\label{f} E[ (X_n/n)^r] \rightarrow \int_\Omega h^r\, d\mu
\q\text{as } n\to\infty.
\end{equation}
Since $h$ is bounded and the above relation holds
for all integers $r\geq 0$, we deduce by the method of moments
the following.
(See \cite{billingsley} for the general theory of weak convergence.)
\begin{theo}\label{T1}
With notation as above,
\begin{equation}
\frac 1nX_n \Rightarrow Y=h(Z)\q\textrm{as } n\to\infty,
\label{clt2}
\end{equation}
where $Z$ is a random element of $\Omega$ with distribution
$\mu$.
\end{theo}
In particular, the support of $Y$ is $[\min h,\max h]$, the
range of $h$, at least provided the density of $\mu$ given above does
not vanish on some interval.

A similar weak limit theorem for $X_n/n$ has been proved
by Konno \cite{konno1, konno2},
by different methods and with a quite different description of the limit.

We note that no assumption has been made above
on the matrix $A$
and the initial state $\psi_0$,
and thus the above result holds for any unitary quantum walk on the integers,
subject only to (\ref{assump}).
Note also that $\mu$ depends only on the overlap
between the initial state of the system and the eigenvectors of
$U(k)$, whereas $h$ depends only on the coin flip matrix $A$.

As an example, we consider some specific cases of unitary quantum
walks. We consider first the Hadamard matrix
\begin{equation}
A= \frac1{\sqrt 2}\left( \begin{array}{cc} 1 & \hfill1 \\ 1 & -1\end{array}
\right).
\end{equation}
By simple calculus,
\begin{equation}
\lambda_j(k)=\frac{i}{\sqrt2} \sin k \pm
\sqrt{1-\tfrac{1}{2}\sin^2 k} \label{lamb}
\end{equation}
and thus
\begin{equation}\label{hkj}
h(k,j)=\frac{-i\lambda_j'(k)}{\lambda_j(k)} =\pm \frac{\cos
k}{\sqrt{2-\sin^2 k}}.
\end{equation}
Hence the limit distribution is concentrated
on the interval
\begin{equation}
[\min h,\max h]=\left[-\frac 1{\sqrt2},\frac 1{\sqrt2}\right].
\label{interval}
\end{equation}

For a general unbiased walk, we take as coin flip the unitary
matrix
\begin{equation}\label{unb}
U(\varphi,\psi)=
 \frac{1}{\sqrt2}\left( \begin{array}{cc} e^{i(\varphi+\psi)}
   & \hfill e^{-i(\varphi-\psi)} \\ e^{i(\varphi-\psi)}
& -e^{-i(\varphi+\psi)}\end{array} \right),
\end{equation}
where $\varphi,\psi \in \RR$, with corresponding evolution
\begin{align}
U(k)&=
 \frac1{\sqrt2}
\left( \begin{array}{cc} e^{i k}e^{i(\varphi+\psi)}
& \phantom{-} e^{i k}e^{-i(\varphi-\psi)} \nonumber\\
e^{-i k}e^{i(\varphi-\psi)}
& - e^{-i k}e^{-i(\varphi+\psi)}\end{array} \right)\\
&\doteq
U_{\varphi \psi}(k).
\end{align}
With $\varphi+\psi=a$, $\varphi-\psi=b$, the eigenvalues
may be written in the form
\begin{equation}
\lambda_j(k)=\frac{i}{\sqrt2} \sin (k+a) \pm
\sqrt{1-\tfrac{1}{2}\sin^2 (k+a)},
\end{equation}
and therefore
\begin{equation}
h(k,j)=\frac{-i\lambda_j'(k)}{\lambda_j(k)} =\pm\frac{ \cos
(k+a)}{\sqrt{2-\sin^2 (k+a)}}.
\end{equation}
Thus the general unbiased walk has exactly the same behaviour
as the Hadamard case,
subject to a shift in the momentum parameter of the wave
amplitudes. We have as before
that the domain of the limit distribution is as in
(\ref{interval}).

Finally we introduce a \lq\lq biased" random walk by defining a
\textit{bias factor} $\rho$ in the coin flip matrix

\begin{equation}
U(\rho)= \left( \begin{array}{cc} \sqrt{\rho} & \sqrt{1-\rho} \\
\sqrt{1-\rho} & -\sqrt{\rho}\end{array} \right),
\end{equation}
that gives rise to the evolution

\begin{equation}
U_{\rho}(k)= \left( \begin{array}{cc} e^{i k}\sqrt{\rho} & e^{i
k}\sqrt{1-\rho}
\\ e^{-i k}\sqrt{1-\rho} & -e^{-i k}\sqrt{\rho}\end{array} \right).
\end{equation}
The evolution under $U$ of a general two-component wave-function
corresponds to
\begin{align}
\left( \begin{array}{c} \psi_1\\ \psi_2\end{array} \right)
\mapsto
&\left( \begin{array}{cc} e^{ik} & 0 \\
0 & e^{-ik}\end{array} \right)\nonumber\\
&\times\left( \begin{array}{c} \sqrt{\rho}
\psi_1(k)+\sqrt{1-\rho}\psi_2(k)\\
\sqrt{1-\rho} \psi_1(k)+\sqrt{\rho}\psi_2(k)\end{array} \right),
\end{align}
where the two internal states transform differently. In fact, the
first component receives a kick of momentum $+k$ with probability
$\rho$ and $-k$ with probability $1-\rho$; the opposite holds
for the second component.

In terms of $\rho$, the eigenvalues are
\begin{equation}
\lambda_j(k)=i \sqrt{\rho} \sin k \pm \sqrt{1-\rho \sin^2 k}
\end{equation}
and thus
\begin{equation}
h(k,j)=\frac{-i\lambda_j'(k)}{\lambda_j(k)}= \pm \frac{\cos
k}{\sqrt{\rho^{-1} - \sin^2k}}.
\end{equation}
It follows that $[\min h,\max h]=[-\sqrt{\rho},\sqrt{\rho}]$, whence
the bias factor of the walk sets a limit on the asymptotic
momentum distribution by changing the support of the limit distribution.

The representation (\ref{clt2}) of the limit variable
allows a direct computation of the asymptotic probability
density function in most cases of interest.
For example, assume that the initial state is at position $0$.
If the coin initially is in
a given state $i=1$ or 2, then $\Psi_0(k)= \smatrixx{1\\0}$
or $\smatrixx{0\\1}$, respectively,
and thus
$\mu=|v_{ji}(k)|^2 \,dk/2\pi$ on $\KK\times\set{j}$.
If we consider instead a random initial state of the coin, we have
a mixture of these two pure states and thus
\begin{equation}
\mu=\frac12\sum_{i=1}^2|v_{ji}(k)|^2 \frac{dk}{2\pi}
=\frac{dk}{4\pi}
\end{equation}
on $\KK\times\set{j}$; that is, $\mu$ is the uniform distribution on $\Omega$.
In the Hadamard case, for example,
with $h$ given by \eqref{hkj},
if $X_0=0$ and the coin initially random,
then, for $-1/\sqrt2 \le y \le 1/\sqrt2$,
\begin{align*}
P(Y\le y)
&=\int_{h^{-1}([-\infty, y])}\,d\mu\\
&=2\int_{\cos k/\sqrt{1+\cos^2 k}\le y}\frac {dk}{4\pi}\\
&= 1-\frac{1}{\pi}\arccos\left(\frac{k}{\sqrt{1-k^2}}\right),
\end{align*}
which gives as density $f(y)$ of $Y$,
\begin{equation}
f(y)=\frac{dy}{\pi (1-y^2)\sqrt{1-2y^2}} ,
\end{equation}
in agreement with the result of \cite{konno1}.
The same holds for every unbiased walk defined by \eqref{unb}.

The above result can be interpreted as the weak
convergence of the sequence $\hat{X}_n/n$ of operators on $H$,
as $n\rightarrow \infty$, to an operator $V$, defined on a dense
subspace of $H$ with spectral resolution
\begin{equation}
V=\int \sum_j \left(\frac{D\lambda_j(k)}{\lambda_j(k)}\right) \,
dE_j(k),
\end{equation}
where $dE_j(k)$ is the projector over the eigenspace corresponding
to the eigenvalue $\lambda_j(k)$ of $U(k)$.
(The weak convergence of unbounded operators here is formally
defined as the weak convergence of the corresponding unitary
operators $\exp(is\hat X_n /n) \to \exp(isV)$ for every real $s$.)
The limit operator is diagonal in the eigenbasis of the unitary
evolution of the walk and gives
\begin{equation}
\langle \hat{X}_n \rangle \sim \langle V \rangle  n,
\end{equation}
that represents the Heisenberg equation of motion for the
position, in the limit $n\rightarrow \infty$, if we interpret $V$
as the \lq\lq velocity" operator. Thus, asymptotically, the centre
of the wave packet moves with constant speed, given by $V$. It is
worth pointing out that, although the equation of motion resembles
the one of a classical system with constant velocity, the state of
the quantum particle spreads in time, with a quadratic growth in
the variance of the position distribution.

%We thus see that the position operators $X_n/n$ are asymptotically
%diagonalized by the vectors $v_j(k)$ in the momentum space.

\section{Weak limit for $d$-dimensional quantum walks}
Let $d\ge 1$.
The classical random walk on the integer lattice $\ZZ^d$ models the
motion of a particle that moves in
an unbiased manner in a $d$-dimensional space.
Let $\ee_i$, $i\in\{1,2,\dots,n\}$, be the unit vector
in the direction of increasing $i$th coordinate. Let $\RRR_1,\RRR_2,\dots$
be independent identically distributed random variables, each being
uniform on the set $\{\pm\ee_i: i=1,2,\dots,d\}$.
The position of the particle at time $n$ is given as
the sum
\begin{equation}
\XX_n=\sum_{j=1}^n\RRR_j.
\end{equation}
By the central limit
theorem for $d$-dimensional random walk, the random vector $\XX_n
/\sqrt{n}$ converges weakly as $n\rightarrow\infty$ to a
random vector in $\RR^d$ having the multivariate normal
distribution $N(0,I/d)$, where $I$ is the $d\times d$ identity matrix.
We shall see
in the following that a corresponding weak convergence holds
for a $d$-dimensional quantum random walk.

The 1-dimensional quantum random walk of the last section
may be extended to $d$ dimensions as follows.
Let $\eps_1,\dots,\eps_{2d}$ denote the $2d$ possible shift vectors
$\pm\ee_i$, $i=1,2,\dots,d$.
The state of the system is a vector
$\Psi=(\Psi({\bf k})_J)_{J=1}^{2d} \in H=
L^2(\KK^d)\otimes \CC^{2d}$ where
${\bf k}=(k_1,k_2,\dots,k_d)$
and the $J$th component corresponds to a shift by the vector
$\eps_J$.
At each time, the state is transformed by
applying a rotation $A$ acting on $\CC^{2d}$, followed by a
$d$-dimensional shift on $L^2(\KK^d)$, cf.\ \eqref{s11},
\begin{equation}
S^{(d)} \Psi({\bf k})_J=e^{i {\bf \eps}_J\cdot {\bf k}} \Psi({\bf k})_J.
\end{equation}
The general unitary operator that
evolves the walk from time $n=0$ is thus
\begin{equation}
U({\bf k})= {\mathcal D}\{e^{i {\bf \eps}_1 \cdot {\bf k}},
\dots, e^{i {\bf \eps}_{2d} \cdot {\bf k}}\} A,
\end{equation}
where ${\mathcal D}$ denotes the $2d$ diagonal matrix. The operator $U({\bf
k})$ can be diagonalized in $H$, and has $2d$ eigenvalues and $2d$
eigenvectors. Assume that one may choose the latter as $C^\infty$
functions of
${\bf k}$.
(See the remark at the end of the section.)
Let $v_J({\bf k})$, $\lambda_J({\bf k})$ be
respectively the eigenvectors and eigenvalues of $U({\bf k})$,
with $J=1,2,\dots, 2d$. The initial state of the system can be
written in this basis as
\begin{equation}
\Psi_0({\bf k})= \sum_{J=1}^{2d}\langle v_J({\bf k}),\Psi_0({\bf
k})\rangle v_J({\bf k})
\end{equation}
and the state at time $n$ as
\begin{equation}
\Psi_n({\bf k})= \sum_{J=1}^{2d} \lambda_J^n({\bf k}) \langle
v_J({\bf k}),\Psi_0({\bf k})\rangle v_J({\bf k}).
\end{equation}

The $d$-dimensional position operator $X^{(d)}=(X_1, X_2,\dots,
X_d)$ acts on $L^2(\KK^d)$ as the differential vector operator
$D^{(d)}=(-id/dk_1, -id/dk_2,\dots, -id/dk_d)$. By considering
each component of $D^{(d)}$ separately, it is easily seen that the
operators $\hat{X}_{i, n}$ converge weakly on $H$, as
$n\rightarrow \infty$, to the corresponding components $V_i$,
where
\begin{equation}
V_i = \int \sum_J \left(\frac{D_i \lambda_J(\bf
{k})}{\lambda_J(\bf {k})}\right)\, {dE_J({\bf k})},
\end{equation}
where $dE_J(\bf {k})$ denotes again the projector onto the eigenspace
of $U({\bf k})$ with eigenvalue $\lambda_J(\bf {k})$. This does
not imply, however, that the sequence of random vectors associated
with the process converges weakly on $\Omega = \KK^d \times
\set{1,\dots,2d}$. In general, the evolution operator $U({\bf
k })$ generates entanglement between the different spatial
directions and it is necessary therefore to consider also the
correlation terms between different components of $\hat{X}^{(d)}$.

The so-called
Cram\'er--Wold device enables a simplification: in order that
a sequence of random variables converge weakly, it suffices that all
linear combinations converge. More properly, we have the
following, see \cite{billingsley}, Theorem 29.4.

\begin{theo}\label{CW}
Consider a sequence
$\XX_n= (X_{1, n},X_{2,n},
\dots, X_{d, n})$, $n\ge 1$, of random $d$-vectors,
and let $\YY=(Y_1,Y_2,\dots,Y_d)$ be a random $d$-vector.
If
\begin{equation}
\sum_{j=1}^d c_j  X_{j, n} \Rightarrow \sum_{j=1}^d c_j  Y_j
\q\text{as } n\to\infty,
\end{equation}
for all $\mathbf{c} = (c_1,c_2,\dots, c_d) \in
\RR^d$,
then $\XX_n\Rightarrow \YY$.
\end{theo}

Suppose for simplicity that $d=2$.
For fixed $r$, we compute
the expectation
\begin{align}
E&\left[\left(\sum_{j=1}^2 c_j  \hat{X}_{j, n}/n\right)^r\right]
\nonumber\\
&= \frac 1{n^r}  \sum_{p=0}^r
\binom{r}{p} c_1^{r-p} \,
c_2^p \; \langle \Psi_n, D_1^{r-p} \, D_2^{p} \,
\Psi_n \rangle,
\end{align}
where we have used the fact that operators along different
directions commute.
% The terms in the sum corresponding to $p=0,
%r$ reduce to (\ref{d}) of the 1-dimensional walk. For all other $p$,
We have
\begin{widetext}
\begin{align}
D_2^p \Psi_n({\bf k})
&= \sum_J (n)_p \lambda_J^{n-p} ({\bf k}) \left( D_2 \lambda_J
({\bf k}) \right)^p \, \langle v_J({\bf k}), \Psi_0({\bf k})
\rangle v_J({\bf k}) + \O (n^{p-1})\\
D_1^{r-p} [D_2^p  \Psi_n({\bf k})]
&= \sum_J (n)_r \lambda_J^{n-r} ({\bf k}) \left( D_1 \lambda_J
({\bf k}) \right)^{r-p} \left( D_2 \lambda_J ({\bf k}) \right)^p
\, \langle v_J({\bf k}), \Psi_0({\bf k}) \rangle v_J({\bf k}) + \O
(n^{r-1}).
\end{align}
\end{widetext}
Thus, as $n \rightarrow \infty$,
\begin{widetext}
\begin{align*}
&E \left[\left( \frac{(c_1 \hat{X}_1 + c_2
\hat{X}_2)_n}{n}\right)^r\right] \\
&\quad\to\int \sum_J \left\{\sum_{p=0}^r \binom rp c_1^{r-p} \, c_2^p
\, h_1 ({\bf k},J)^{r-p} h_2 ({\bf k},J)
^p\right\}|\langle v_J({\bf k}), \Psi_0({\bf k}) \rangle|^2 \frac{d{\bf
k}}{(2\pi)^2}\\
&\quad=\int \sum \{ c_1 h_1 ({\bf k},J) + c_2 h_2 ({\bf k},J)\}^r \,
|\langle v_J({\bf k}), \Psi_0({\bf k}) \rangle|^2 \frac{d{\bf
k}}{(2\pi)^2} ,
\end{align*}
\end{widetext}
where $h_i({\bf k},J)=\lambda_J ({\bf k})^{-1}D_i \lambda_J ({\bf k})$,
$i=1, 2$. With $\Omega = {\KK}^2 \times \{1, 2, 3, 4\}$,
and $Z_n = c_1 X_{1,n} + c_2 X_{2,n} $, we
have
\begin{equation}
E\left[({Z_n}/{n})^r\right] \rightarrow \int_{\Omega} (c_1h_1+c_2h_2)^r \,
d\mu,
\end{equation}
where $\mu$ is the probability measure on $\Omega$ given by
\begin{equation}
  \label{mud}
d\mu=
|\langle v_J({\bf k}), \Psi_0({\bf k}) \rangle|^2 \frac{d{\bf
k}}{(2\pi)^2} \quad \text{on } {\KK}^2 \times \{J\}.
\end{equation}
By the method of moments as in the one-dimensional case and the
Cram\'er--Wold device (Theorem \ref{CW}), we obtain a
generalization of Theorem \ref{T1} to the two-dimensional case.

As a simple example, consider the two-dimensional generalization
of the Hadamard matrix given by
\begin{equation}
A= \frac1{ 2}\left( \begin{array}{rrrr} 1 & 1 & 1 & 1\\ 1 & -1 & 1
& -1\\1 & 1 & -1 & -1\\1 & -1 & -1 & 1
\end{array} \right).
\end{equation}
In the above notation, the unitary operator that evolves the walk
is represented by
\begin{equation}
U({\bf k})= {\mathcal D}\{e^{i k_1}, e^{i k_2}, e^{-i k_2}, e^{-i
k_1}\} A.
\end{equation}
The operator
$U({\bf k})$ may be expressed thus as a tensor product of two
one-dimensional operators that describe Hadamard walks along the
directions defined by $k^+ = (k_1 + k_2)/2$ and $k^- = (k_1 -k_2)/2$:
\begin{equation}
U({\bf k})= U(k^+) \otimes U(k^-).
\end{equation}
Its eigenvalues and eigenvectors are products of those of
$U(k^+)$,  $U(k^-)$ respectively, and therefore
\begin{equation}
\lambda_J ({\bf k}) = \lambda_{j_+}(k^+) \, \lambda_{j_-}(k^-),
\end{equation}
where the $\lambda_{j}$, $j= 1, 2$, are given by (\ref{lamb}) and
$J=1,2,3,4$ labels the pairs $(j_+,j_-)$ in some order. Thus
\begin{align}\label{bigh}
h_i({\bf k},J)
={}&\lambda_J ({\bf k})^{-1}D_i \lambda_J ({\bf k})\nonumber\\
={}&\pm \frac{\cos (k^+)}{2 \sqrt{2-\sin^2 (k^+)}} \nonumber\\
&\pm (-1)^{i-1}\frac{\cos (k^-)}{2 \sqrt{2-\sin^2 (k^-)}}.
\end{align}
for $i=1,2$.
The limit velocity operator ${\bf V}$ is given by ${\bf V}= (V_1,
V_2)$, with
\begin{equation}
V_i = \int \sum_J h_i({\bf k},J) \, {dE_J({\bf k})},
\quad J=1, 2, 3, 4.
\end{equation}

The result may be extended to arbitrary dimension $d \ge 2$ using
the
same argument, yielding the following result.
\begin{theo}\label{Td}
For the $d$-dimensional quantum random walk,
\begin{equation}
\frac 1n \XX_n \Rightarrow \YY=\bigl(h_1(Z),\dots,h_{d}(Z)\bigr),
\label{clt2d}
\end{equation}
where
$Z$ is a random element of
$\Omega = {\KK}^2 \times \{1, \dots, 2d\}$
with distribution
$\mu$ given by \eqref{mud}
and
$h_i({\bf k},J)=\lambda_J ({\bf k})^{-1}D_i \lambda_J ({\bf k})$.
\end{theo}

The limit
observable is again diagonal in the eigenbasis of $U(\bf{k})$ and
represents the velocity for
$n\rightarrow \infty$.

\begin{remark}
We assumed above that the eigenvectors of
$U({\bf k})$ can be chosen
as $C^\infty$
functions of
${\bf k}$. We do not know if this is always possible when $d\ge2$, but
it can be replaced by the following, weaker hypothesis, which we
believe always holds:
There exists an
open subset
$O$ of $\KK^d$ with full Lebesgue measure (that is,
the complement is a null set)
such
that the eigenvectors (and thus the eigenvalues)  can be chosen
infinitely differentiable in $O$.
(For example, this holds if there is any point
${\bf k}$
where $U({\bf k})$ has distinct eigenvalues, because we then can choose
$O$ as the subset of $(0,2\pi)^d$ where the discriminant is non-zero;
we omit the proof that this set has the required properties.)

Under this assumption, the argument above
holds for every initial value that is an
infinitely differentiable function with support in $O$.
(The function
$h$ will be defined on $O\times\{1,2,\dots,2d\}$, but that is enough.)
Such functions are dense in $H$, by a standard $L^2$ result.
Hence, given any initial state $\Psi_0$, and an $\epsilon>0$,
we can find an initial state $\Psi^\epsilon_0$ with
$\|\Psi_0-\Psi^\epsilon_0\|<\epsilon$ for which
\begin{equation}\label{ceps}
\frac 1n \XX_n^\epsilon
\Rightarrow \bigl(h_1(Z^\epsilon),\dots,h_{d}(Z^\epsilon)\bigr).
\end{equation}
Since the evolution operators are unitary, we have
$\|\Psi_n-\Psi^\epsilon_n\|
=\|\Psi_0-\Psi^\epsilon_0\|<\epsilon$ for every $n$, and it follows that
for any observable event $A$,
the probabilities $\PP(\XX_n \in A)$ and $\PP(\XX^\epsilon_n \in A)$
differ by at most $2\epsilon$.
Similarly, it is easy to see from
\eqref{mud} that $|\PP(Z \in B)-\PP(Z^\epsilon \in B)|<2\epsilon$
for every $B\subset\Omega$.
It is now easy to
interchange the two limits $\epsilon\to0$ and $n\to\infty$
and obtain \eqref{clt2d};
see \cite{billingsley}, Theorem 4.2.
Theorem \ref{Td} thus holds for
every initial state, also under the weaker assumption.
\end{remark}

\section{Further extensions}
We have, for simplicity, only considered simple random walks, where
the shifts are by unit vectors. More generally, we can allow shifts by
any given finite set $\set{\eps_1,\eps_2,\dots,\eps_N}$ of vectors
in $\mathbb Z^d$. The coin flip is now represented by a unitary matrix
$A$ in $\mathbb C^N$. Theorem \ref{Td} extends to this case,
with $2d$ replaced by $N$, by the same proof.

An interesting example is when the shift vectors are the $2^d$ vectors
in $\set{-1,1}^d$; thus each coordinate is shifted by $\pm1$ in each step.

\acknowledgments

PFS would like to thank the Statistical Laboratory, University of
Cambridge, for its warm hospitality. Work by PFS was supported by
the British Technion Society. This paper was completed during a
programme at the Isaac Newton Institute, Cambridge.

\end{document}